\documentclass[twocolumn,aps,prl,superscriptaddress]{revtex4-1}

\usepackage{slashed}
\usepackage{enumitem}

\usepackage{float}

\usepackage{tikz-cd}
	\usepackage{amsmath}
	\usepackage{amsfonts}
	\usepackage{amssymb}
	\usepackage{graphicx}
	\usepackage{mathtools}
	\usepackage{stmaryrd}
	\usepackage{autobreak} 
    \usepackage{cancel}
 
	\allowdisplaybreaks
	\usepackage[colorlinks=true,citecolor=blue,linkcolor=red]{hyperref}
	\usepackage{bbold}					
	\usepackage{multirow}				
	\usepackage[normalem]{ulem}        
	\usepackage{array} 
    \usepackage{cancel} 
\usepackage{lipsum}
\usepackage{graphicx}

\usepackage{titletoc} 
\usepackage{mathtools}

	\newcolumntype{x}[1]{>{\centering\let\newline\\\arraybackslash\hspace{0pt}}p{#1}}
	
	\makeatletter
\newcommand*\rel@kern[1]{\kern#1\dimexpr\macc@kerna}
\newcommand*\widebar[1]{%
  \begingroup
  \def\mathaccent##1##2{%
    \rel@kern{0.8}%
    \overline{\rel@kern{-0.8}\macc@nucleus\rel@kern{0.2}}%
    \rel@kern{-0.2}%
  }%
  \macc@depth\@ne
  \let\math@bgroup\@empty \let\math@egroup\macc@set@skewchar
  \mathsurround\z@ \frozen@everymath{\mathgroup\macc@group\relax}%
  \macc@set@skewchar\relax
  \let\mathaccentV\macc@nested@a
  \macc@nested@a\relax111{#1}%
  \endgroup
}
\makeatother

	\DeclareMathAlphabet{\mathbbold}{U}{bbold}{m}{n}

	\newcounter{subeqn} %
	\makeatletter
	\@addtoreset{subeqn}{equation}
	\makeatother

\definecolor{ZM}{rgb}{.5,0,.5}
\definecolor{SD}{rgb}{0,1,0}

\newcommand\trick[1]{} 

\begin{document}
\title{Far-from-equilibrium topological phase transition in one dimension }

\author{Ze-Min Huang}
\affiliation{Institute for Theoretical Physics, University of Cologne, 50937 Cologne, Germany}

\author{Gustav John}
\affiliation{Institute for Theoretical Physics, University of Cologne, 50937 Cologne, Germany}

\author{Sebastian Diehl}
\affiliation{Institute for Theoretical Physics, University of Cologne, 50937 Cologne, Germany}

\begin{abstract}
We uncover a mechanism for far-from-equilibrium topological phase transitions, via a one-dimensional compact phase model evolving deterministically from random initial conditions. It rests on topology and symmetry rather than on phenomenological postulates: phase compactness permits vortices, and a homogeneous fixed point suppresses their nucleation, with the fixed point itself implied by phase-shift symmetry.
The competition between vortex-induced disordering and relaxation toward homogeneity drives a continuous nonequilibrium transition, whose universality class we identify as directed percolation (DP). We demonstrate this by constructing the corresponding effective field theory and numerically confirming DP critical scaling through dynamical-scaling analysis.
\end{abstract}

\maketitle

\textit{\color{red}{Introduction.--}} The proliferation of topological defects underlies some of the most striking universal phenomena in equilibrium physics, from the Berezinskii-Kosterlitz-Thouless (BKT) transition in condensed matter~\cite{berezinskii1971jetp,kosterlitz1973jpc} to confinement in gauge theories~\cite{polyakov1987harwood}. Far from equilibrium, defects are no less pervasive~\cite{aranson2002rmp,shankar2022nrp,sieberer2025rmp,fruchart2026arxiv}, as topology determines which defects can exist, while dynamics determines whether they decay, bind, proliferate, or even support spatiotemporal chaos~\cite{hecke2001prl}.
Yet the universal consequences of the nonequilibrium dynamics remain poorly understood.
For spatial defects defined on an equal-time slice, a paradigmatic example is the one-dimensional directed-Ising universality class, governed by branching and pair-annihilating domain walls~\cite{cardy1996prl,cardy1998jsp,tauber2014cambridge}. Much less is known about topological defects localized in spacetime, instanton events that connect configurations across time, such as spacetime vortices in $(1+1)$ dimensions~\cite{aranson2002rmp,sieberer2025rmp}
and monopole events in $(2+1)$ dimensions. Whether such spacetime defects can drive continuous criticality, and if so, through what mechanism and in which universality class remains unresolved.

\begin{figure}

\includegraphics[width=1\linewidth,keepaspectratio]
{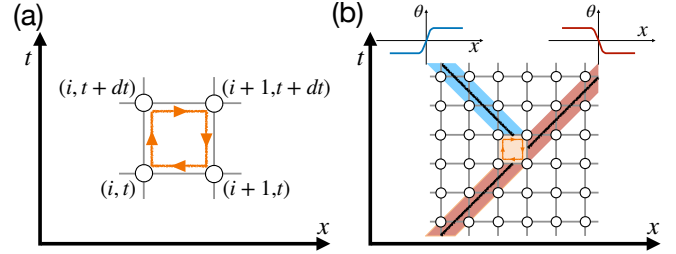}
\centering
\caption{Schematic illustration of (a) the definition of a spacetime vortex and (b) its multiplicative character and disordering effect on the phase field. (a) The vortex charge on a plaquette is defined as the oriented sum of the link field $[\mathcal{D}_\mu\theta]_{2\pi}$ along its boundary, where $\mu=x,t$ and $[\cdots]_{2\pi}$ denotes reduction to the principal branch $(-\pi,\pi]$. (b) A spacetime vortex can occur only when the phase field is inhomogeneous; across the event, the local spatial winding carried by a kink changes by the vortex charge, enabling kink branching.\label{fig:illustration}}

\end{figure}

In this Letter, we address these questions in one spatial dimension using a compact phase model with deterministic discrete-time dynamics and a homogeneous fixed point. The compactness of the phase field crucially permits spacetime vortices [Fig.~\ref{fig:illustration}(a)], whose singular core underlies the physical origin of the temporal discreteness. These vortex events promote disorder [Fig.~\ref{fig:illustration}(b)]: when a kink, a localized region of large phase gradient, encounters a spacetime vortex, it can generate additional kinks, leading to a branching process. By contrast, relaxation toward the homogeneous fixed point favors an ordered phase configuration. The competition between these mechanisms drives a continuous nonequilibrium phase transition, which we show belongs to the DP universality class. The critical scaling of DP is known to hinge on multiplicative noise~\cite{hinrichsen2000sa}, whereas the present model is fully deterministic. We reconcile these facts by showing that spacetime vortices, acting as fast, singular degrees of freedom, generate effective noise for the underlying slow mode when integrated out. This noise is necessarily multiplicative [Fig.~\ref{fig:illustration}(b)], since fluctuations and thus vortex production vanish at the homogeneous fixed point. Guided by symmetry, we further construct a Landau-Ginzburg-Wilson effective theory, and show that it reduces to the Reggeon field theory of directed percolation~\cite{hinrichsen2000sa,janssen2005aop}, consistent with the dynamical scaling of our numerical data.

\textit{\color{red}{From vortices to multiplicative noise.--}}
Spacetime vortices are topological defects whose definition is independent of the microscopic dynamics. We therefore study their proliferation in a generic one-dimensional lattice of $N$ sites,
\begin{equation}
\mathcal{D}_{t}\theta\left(i,t\right)=\mathcal{K}_{i}\left[\mathcal{D}_{x}\theta\right].\label{eq:eom_phase}
\end{equation}
Here, the lattice spacing is set to unity $d x=1$, and   $\mathcal{D}_x\theta(i, t)\equiv\left[\theta\left(i+1,t\right)-\theta\left(i,t\right)\right]$ denotes the lattice difference with $i=1,2 \dots $ labeling the lattice sites.  $\mathcal{K}_{i}$ is a bounded local map associated with $\theta$ at site $i$, subject to the following constraints from topology and symmetry:  (i) \textit{compactness}, i.e., $2\pi$ periodicity in its argument, consistent with the compact phase field $\theta\simeq\theta+2\pi$; (ii) the \textit{fixed-point condition}, $\mathcal{K}_{i}[\mathcal{D}_x\theta =0]=0$ admitting a spatially homogeneous fixed point; and (iii) \textit{spatial inversion symmetry}, $P\theta(i, t)=\theta(N-i+1, t)$ and $P\mathcal{K}_
{i}=\mathcal{K}_
{N-i+1}$, imposed for simplicity to preserve vortex neutrality.
Crucially, the fixed-point condition here is not an independent dynamical assumption for an absorbing state, but a direct consequence of the symmetry of a phase, namely, of
global phase-shift symmetry after removing any uniform rotation. Physically, such a phase-only description naturally arises from an underlying complex-field model, e.g., a Gross–Pitaevskii model (see below for a numerical illustration), when amplitude fluctuations are suppressed relative to phase fluctuations, leaving the phase as the dominant mode with global phase-shift symmetry. This reduction is well defined away from spacetime-vortex cores, where the amplitude vanishes and the phase becomes ill-defined. The resulting phase singularity appears as a discontinuous phase jump, which can be retained within a phase-only description through temporal discretization, thereby motivating the discrete-time dynamics considered here. 
Finally, the specific form of $\mathcal{K}_{i}$, introduced below, is unimportant
for the current discussion. 

Importantly, despite the deterministic dynamics of Eq.~\eqref{eq:eom_phase}, the slow modes, whose dynamics is effectively continuous in time, experience noise generated by spacetime vortices. In particular, the microscopic dynamics enters this continuous-time description in two distinct ways. (i) Away from singular vortex events, the limit $dt\to0^{+}$ may be taken directly,
yielding the regular continuous-time dynamics. (ii) At a vortex event, however, this naive limit fails. The associated topological obstruction is captured by an effective noncommutativity of the temporal and spatial derivatives. Symbolically, a vortex of charge $q_{v}$ located at $\left(i_{v},t_{v}\right)$
contributes (see below for a precise lattice formulation)
\begin{equation}\label{eq:top_obstruction}
\left(\partial_{t}\mathcal{D}_{x}-\mathcal{D}_{x}\partial_{t}\right)\theta\left(i,t\right)=j_{v},\ \mathrm{and}\ j_{v}=2\pi q_{v}\delta\left(t-t_{v}\right)\delta_{i,i_{v}},
\end{equation}
where $j_{v}$ is the vortex charge density, and $\partial_{t}$ denotes the continuous-time counterpart of $\mathcal{D}_{t}$. Since $\theta$ is singular at vortex events, it is the phase gradient that provides the natural slow field,
\begin{equation}
u\left(i,t\right)\equiv\mathcal{D}_{x}\theta\left(i,t\right),\ \text{and}\  \partial_{t}u\left(i,t\right)=\mathcal{D}_{x}\mathcal{K}_{i}\left[u\right]+j_{v},\label{eq:slow_mode_vortex}
\end{equation}
with $u\in(-\pi, \pi]$ and the second relation following from Eq.~\eqref{eq:top_obstruction}. Upon coarse-graining, vortex events act as an effective noise source, as we show below.

\begin{figure}
\includegraphics[width=1\linewidth,keepaspectratio]{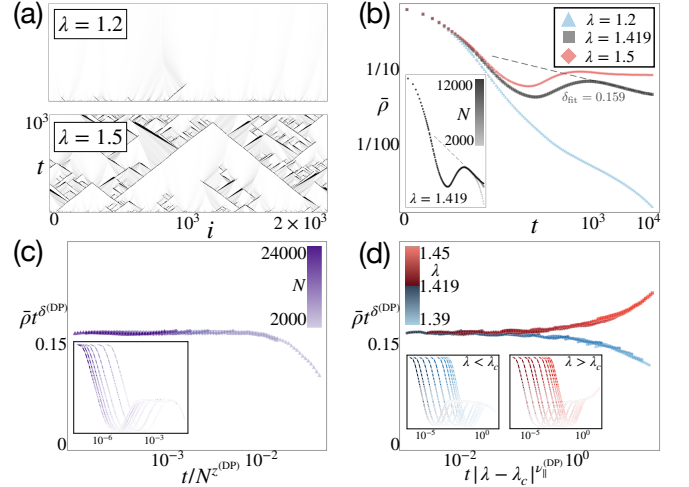}
\centering
\caption{Numerical results for the discrete KPZ model using $D=\frac{1}{2}$ and $dt=\frac{1}{4}$.
(a) Spacetime profiles of the local observable $\rho(i, t)$.
At late times, the system either approaches the $u=0$ state,
or remains spatially inhomogeneous, with branching structures reminiscent
of DP. Panels (b)-(d) are averaged over $10^{4}$
random initial conditions. (b) Log-log plot of $\bar{\rho}(t)$.
For $\lambda\ll\lambda_{c}$, $\bar\rho$ decays rapidly, whereas
for $\lambda\gg\lambda_{c}$ it approaches a long-lived active state.
At $\lambda_{c}\simeq1.419$, $\bar{\rho}$ exhibits algebraic
decay with fitted exponent $\delta_{\mathrm{fit}}=0.159$ (over $t\in\left[10^{3},10^{4}\right]$),
consistent with the DP value $\delta^{\left(\mathrm{DP}\right)}=0.159464$.
The inset shows the system-size dependence at $\lambda=\lambda_{c}$.
(c) Finite-size scaling collapse over the same time window, using
the DP critical exponent, $z^{\left(\mathrm{DP}\right)}=1.580745$.
(d) Off-critical scaling collapses for $N=8000$ using $\nu^{\left(\text{DP}\right)}_{\parallel}=1.733847$.
The insets in (c,d) show the full time range. The fitting window is
chosen such that $t/N^{z^{\left(\text{DP}\right)}}\ll1$, thereby
taking the thermodynamic limit before the long-time limit.
This order of limits defines DP criticality, necessitating a dynamical determination of the critical exponents, since the true steady state is not directly accessible numerically~\cite{hinrichsen2000sa}.
\label{fig:discrete_KPZ_snapshot}}
\end{figure}

The vortex-induced noise is necessarily multiplicative because Eq.~\eqref{eq:eom_phase} admits a homogeneous fixed point at which vortices cannot nucleate. To make this precise, we first define vortices on the spacetime lattice and then trace them out.  Specifically, a spacetime vortex is naturally associated with an elementary plaquette, as the
phase is defined on spacetime lattice sites. Its charge is then the oriented sum of the phase increments $\mathcal{D}_{\mu}\theta$ around the plaquette boundary [see Fig.~\ref{fig:illustration}(a)], 
\begin{equation}
\begin{aligned}
2\pi q_v(i,t)
={}& \left[dx\mathcal{D}_x\theta(i,t+dt)\right]_{2\pi}+\left[dt\mathcal{D}_t\theta(i,t)\right]_{2\pi} \\
 &+ \left[-dx\mathcal{D}_x\theta(i,t)\right]_{2\pi}
   + \left[-dt\mathcal{D}_t\theta(i+1,t)\right]_{2\pi},
\end{aligned}
\label{eq:vortex_charge}
\end{equation}
where $\left[\dots\right]_{2\pi}$ denotes reduction to a principal
branch $(-\pi,\ \pi]$, and hereafter,  "vortex" denotes a spacetime vortex, but not a vortex in a fixed-time spatial configuration.
We then remove the explicit principal-branch reductions $[\dots]_{2\pi}$ by introducing integer-valued link fields $m_{\mu}(i,t)\in\mathbb{Z}$, with $\mu=x,t$, that isolate the winding contribution to each phase increment, i.e., 
\begin{equation}
dx^{\mu}\mathcal{D}_{\mu}\theta=\left[dx^{\mu}\mathcal{D}_{\mu}\theta\right]_{2\pi}+2\pi m_{\mu}\ \left(\text{no sum over \ensuremath{\mu}}\right),\label{eq:gauge_field}
\end{equation}
which reduces the vortex charge Eq.~\eqref{eq:vortex_charge} to the lattice curl of $m_{\mu}$, 
\begin{equation}
q_{v}=-dt\mathcal{D}_{t}m_{x}\left(i,t\right)+dx\mathcal{D}_{x}m_{t}\left(i,t\right).
\end{equation}
Clearly, this charge vanishes for a spatially homogeneous configuration,
$\mathcal{D}_{x}\theta(i,t)=0$: Homogeneity implies $m_x(i,t)=0$, and since $\mathcal{K}_{i}[0]=0$, the dynamics preserves this condition, so $m_x(i,t+dt)=m_t(i,t)=0$. Therefore, $q_v(i,t)=0$, and vortices cannot dynamically nucleate from a homogeneous configuration. Together with Eq.~\eqref{eq:slow_mode_vortex}, this establishes the multiplicative character of the vortex-induced noise. Indeed, consider an ensemble of trajectories generated by evolving random initial conditions under the deterministic dynamics Eq.~\eqref{eq:eom_phase}. Upon integrating out the vortices, their cumulative effect appears as an effective stochastic force on the slow field $u$. Its local strength is set by the vortex creation rate, denoted by the fugacity $\kappa(u)$, which vanishes in the homogeneous state,
\begin{equation}\label{eq:kappa_u0}
u=0 \implies \kappa(u)=0,
\end{equation}
exhibiting field dependence and thus making the noise multiplicative.
Combined with locality and the absence of internal symmetries, this structure points to the DP universality class~\cite{janssen1981zpc,grassberger1982zpb}, as derived below from the coarse-grained theory and confirmed numerically in
Fig.~\ref{fig:discrete_KPZ_snapshot}.

\textit{\color{red}{Multiplicative noise from the Martin-Siggia-Rose framework.--}} Quantitatively, we integrate out the vortices in the dilute regime relevant near the transition. There, the slow field is nearly homogeneous over most of spacetime, so the associated coarse-grained fugacity is small. We therefore adopt a vortex-gas approximation: the fast events are treated as independent for a given slow-field configuration, while their coupling through the slow field is retained in the local fugacity $\kappa(u)$. This construction is analogous to the vortex-gas description of the BKT transition. Here, however, the fugacity is dynamically slaved to the local phase inhomogeneity and vanishes in the homogeneous state. 

We now implement this systematically using the Martin-Siggia-Rose (MSR) formalism, starting from a fixed vortex configuration $j_{v}$. The MSR partition function is, 
\begin{eqnarray}
Z_{v}\left[j_{v}\right] & = & \int\mathcal{D}u\delta\left(\partial_{t}u-\mathcal{D}_{x}\mathcal{K}_{i}-j_{v}\right)\nonumber \\
 & = & \int\mathcal{D}u\mathcal{D}\tilde{\phi}e^{iS_{0}+iS_{v}\left[j_{v}\right]},
\end{eqnarray}
where $\tilde{\phi}$ is the MSR response field, and 
\begin{equation}
S_{0}=\sum_{i}\int dt\tilde{\phi}\left(\partial_{t}u-\mathcal{D}_{x}\mathcal{K}_{i}\right),\ S_{v}\left[j_{v}\right]=-\sum_{i}\int dt\tilde{\phi}j_{v}.
\end{equation}
We then average over
vortex configurations using the approximation discussed above, treating them as collections of independent, localized elementary vortices with charges $q^{\left(e\right)}_{v}=\pm1$, where the superscript $(e)$ denotes the elementary contribution. Accordingly, a vortex at $\left(i_{v},t_{v}\right)$
is represented by $j^{\left(e\right)}_{v}\left(i,t\right)=\pm2\pi\delta\left(t-t_{v}\right)\delta_{i,i_{v}}$, and carries the MSR weight $e^{iS_{v}\left[j^{\left(e\right)}_{v}\right]}=e^{\mp i2\pi\tilde{\phi}\left(i_{v},t_{v}\right)}$. Taking the same local occurrence rate $\kappa(u)$ for both charge sectors, as required by inversion symmetry, and assuming independent Poisson statistics~\cite{fn1_instanton}, the local average over $n_q$ vortices of charge $q_v^{(e)}=\pm 1$ is \begin{equation}
\sum_{n_q=0}^{\infty} P_{\kappa}(n_q) \left[e^{-i2\pi q_v^{(e)}\tilde\phi}\right]^{n_q}= \exp\left\{\kappa dt\left[e^{-i2\pi q_v^{(e)}\tilde\phi}-1\right] \right\},
\end{equation}
with the Poisson probability $P_{\kappa}(n_{q})\equiv e^{-\kappa(u) dt}[\kappa(u) dt]^{n_{q}}/n_{q}!$.
Multiplying over the two sectors and all spacetime points then gives the vortex-averaged partition function,
\begin{equation}
Z\!=\!\int\!\mathcal{D}u\,\mathcal{D}\tilde\phi\,
\exp\!\left\{iS_0+2\sum_i\int\!dt\,\kappa(u)
[\cos(2\pi\tilde\phi)-1]\right\}.
\label{eq:averaged_partition}
\end{equation}

This partition function makes the stochastic interpretation explicit. Expanding about the physical MSR saddle $\tilde{\phi}=0$ to quadratic order gives $-4\pi^{2}\sum_{i}\int dt\kappa\left(u\right)\tilde{\phi}^{2}+\mathcal{O}\left(\tilde{\phi}^{4}\right)$, which maps to a Gaussian noise theory via a standard Hubbard-Stratonovich transformation, 
\begin{equation}
Z=\int\mathcal{D}\xi P\left[\xi\right]\int\mathcal{D}u\mathcal{D}\tilde{\phi}e^{i\sum_{i}\int dt\tilde{\phi}\left(\partial_{t}u-\mathcal{D}_{x}\mathcal{K}_{i}-\xi\right)},\label{eq:msr_noise}
\end{equation}
with $P\left[\xi\right]=e^{-\sum_{i}\int dt\frac{1}{\left(4\pi\right)^{2}\kappa\left(u\right)}\xi^{2}}$, yielding a Langevin theory with multiplicative noise set by the local vortex fugacity $\kappa(u)$,
\begin{equation}\label{eq:noise_vortex}
\langle\xi\left(i,t\right)\xi\left(i^{\prime},\ t^{\prime}\right)\rangle=8\pi^{2}\kappa\left(u\right)\delta\left(t-t^{\prime}\right)\delta_{i,i^{\prime}}.
\end{equation}
Meanwhile, the higher-order terms in $\tilde{\phi}$ encode higher local noise cumulants, which are renormalization-group (RG) irrelevant for the models considered below~\cite{janssen2005aop}. In turn, this suggests that the universal long-wavelength behavior is insensitive to the detailed form of the vortex averaging, provided that it preserves the same symmetries and the absorbing-state constraint Eq.~\eqref{eq:kappa_u0}. We demonstrate this below by constructing the corresponding Landau–Ginzburg–Wilson theory.

\textit{\color{red}{Directed percolation from vortex proliferation.--}} The mechanism established above hinges solely on symmetry, topology, and a homogeneous fixed point. It  assumes that vortices can be generated dynamically, but without specifying the underlying dynamics. We now exemplify this mechanism and the vortex creation in a concrete model that satisfies Eq.~\eqref{eq:eom_phase} and features competing homogenizing and disordering tendencies, i.e., 
\begin{eqnarray}
\mathcal{K}_{i} & = & D\left[\sin\left(u_{i}\right)-\sin\left(u_{i-1}\right)\right]\nonumber \\
 &  & +\lambda\left[\sin^{2}\left(\tfrac{1}{2}u_{i}\right)+\sin^{2}\left(\tfrac{1}{2}u_{i-1}\right)\right], \label{eq:dKPZ}
\end{eqnarray}
with $u(i, t)\equiv \mathcal{D}_x \theta(i, t)$. 
The $D$ term generates a conservative force that homogenizes the phase field through diffusion; for small gradients, it reduces to the discrete Laplacian $\mathcal{D}_x^2\theta(i,t)$. By contrast, the nonequilibrium $\lambda$ term favors rapid phase variations and, for small $u$, reduces to a term proportional to $[\mathcal{D}_x\theta(i,t)]^2$, the Kardar-Parisi-Zhang (KPZ) nonlinearity~(see e.g., Refs.~\cite{sakaguchi1988ptp,kim2004prb, altman2015prx, he2015prb,he2017prl, lauter2017pre} for the continuous-time case). We therefore refer to Eq.~\eqref{eq:dKPZ} as the discrete KPZ model. However, at finite $\lambda$, including near the transition, numerical results reveal sharp, localized phase variations that invalidate the smooth-gradient expansion, a crucial breakdown that enables vortex generation and places the system far outside the smooth KPZ regime. Additionally, the dynamics preserve the topological winding density $w(t)\equiv \frac{1}{N}\sum_i u(i, t)$. For the random initial ensemble considered here, both the mean and variance of $w(0)$ vanish in the thermodynamic limit,
so the ensemble concentrates in the zero-winding-density sector, which is preserved by the dynamics. Within this sector, $u=0$ is the relevant homogeneous absorbing state, while $w$ is a fixed label rather than a fluctuating mode.

The model undergoes a transition from a phase-coherent regime at $\lambda/D\ll1$ to an inhomogeneous regime at $\lambda/D\gg1$, as illustrated in Fig.~\ref{fig:discrete_KPZ_snapshot}(a). Specifically, we quantify the inhomogeneity by 
\begin{equation}\label{eq:local_order_rho}
\rho(i,t)\equiv \sin^{2}\left[\tfrac{1}{2}u(i,t)\right],\ \bar{\rho}(t)\equiv \frac{1}{N}\sum_i\rho(i,t),
\end{equation}
the simplest measure that is nonnegative, $2\pi$-periodic and vanishes in the homogeneous configuration. For $\lambda/D\ll1$, the homogeneous configuration $u=0$ is linearly stable for $0\leq Ddt\leq1/2$, implying $\bar{\rho}=0$; see Appendix~\hyperref[sec:Linear_stability_anlysis]{A}. At the opposite endpoint $D=0$, an even-site chain acquires the accidental symmetry
$u(i, t)\rightarrow u(N-i, t)+(-1)^i\pi$,
under which $\bar{\rho}\rightarrow1-\bar{\rho}$. For the random initial ensemble, which is invariant under this transformation, the ensemble-averaged order parameter is therefore fixed to $\bar{\rho}=1/2$.
Numerically, we verify this scenario using the standard dynamical-scaling protocol~\cite{hinrichsen2000sa}. We locate the critical point from the algebraic decay $\bar{\rho}(t)\sim t^{-\delta}$ [Fig.~\ref{fig:discrete_KPZ_snapshot}(b)], while the finite-size and off-critical collapses in Figs.~\ref{fig:discrete_KPZ_snapshot}(c) and \ref{fig:discrete_KPZ_snapshot}(d) agree with the one-dimensional DP exponents. We further numerically support vortex proliferation as the physical origin of disorder by showing that the spacetime vortex density grows in parallel with $\bar{\rho}$ in Appendix~\hyperref[sec:rho_vortex]{B}.

\textit{\color{red}{Effective theory for the phase transition.--}}
Having identified vortices as the microscopic source of multiplicative noise, we now construct the long-wavelength Landau-Ginzburg-Wilson theory and show that this noise is RG relevant. As usual, we organize the effective theory in powers of (i) the slowly varying relevant order field, (ii) gradients, and (iii) the distance from criticality. Here, no internal symmetry requires an additional critical field, so the relevant field is the coarse-grained  $\rho(x,t)\equiv\frac{1}{N_{\mathcal{B}}}
\sum_{i\in\mathcal{B}_x}
\sin^2\left[\frac{u(i,t)}{2}\right]$ averaged over spatial blocks $\mathcal{B}_x$ containing $N_{\mathcal{B}}\gg1$ sites, using the same symbol as for the microscopic fields Eq.~\eqref{eq:local_order_rho} when unambiguous.
Near the transition, coarse graining renders $\rho$ slowly varying and small because regions with finite $u$ are dilute, while $u$ remains finite locally. Consequently, a small-$u$ expansion of Eq.~\eqref{eq:noise_vortex} misses the relevant noise scaling, whereas an expansion in $\rho$ yields a linear noise variance at leading order, characteristic of DP~\cite{munoz2003prl}. Moreover, the transition has the standard ingredients associated with DP universality~\cite{janssen1981zpc,grassberger1982zpb}: (i) a continuous active-to-absorbing transition characterized by a scalar order parameter, (ii) short-range dynamical rules, and (iii) no additional fluctuating conserved field or internal symmetry. Although the microscopic absorbing configuration is not unique, the presence of multiple absorbing configurations does not necessarily preclude DP scaling~\cite{jensen1993prl,jensen1994jpa,albano1995pasma,munoz1996prl,hinrichsen2000sa,deger2022prl}. 

We now make this connection precise. Taking the continuous-time, low-frequency limit of Eq.~\eqref{eq:dKPZ} yields the Langevin description implied by Eq.~\eqref{eq:msr_noise}, whose DP critical scaling is confirmed numerically in Appendix~\hyperref[sec:cKPZ_numerics]{C}. Tuning $\lambda$ with $D$ fixed for simplicity (e.g., $D=\frac{1}{2}$), the leading coarse-grained theory is
\begin{equation}
\begin{cases}
\partial_{t}\rho=D_{\text{eff}}\partial^{2}_{x}\rho+m_{\text{eff}}\left(\lambda-\lambda_{c}\right)\rho-g_{\text{eff}}\rho^{2}+\zeta\\
\langle\zeta\left(x,t\right)\zeta\left(x^{\prime},t^{\prime}\right)\rangle=\sigma_{\text{eff}}\rho\left(x,t\right)\delta\left(t-t^\prime\right)\delta\left(x-x^\prime\right)
\end{cases},\label{eq:reggon_theory}
\end{equation}
with $\partial_x$ the continuum counterpart of $\mathcal{D}_x$, $\partial_x^2\rho$ the leading gradient term allowed by inversion symmetry, and $m_{\mathrm{eff}}, g_{\mathrm{eff}}>0$ fixed by the limiting behavior at small and large $\lambda/D$.
Expanding the noise variance in $\rho$, the absorbing-state constraint forbids a constant term, generically leaving $\sigma_{\mathrm{eff}}\rho +\mathcal{O}(\rho^2)$; a leading $\rho^2$ term would instead require fine tuning~\cite{janssen2005aop}, consistent with Appendix~\hyperref[sec:rho_vortex]{B}. Equation~\eqref{eq:reggon_theory} is thus the Langevin representation of Reggeon field theory, dictated by locality, symmetry, and the absorbing-state constraint, thereby establishing the robustness of long-wavelength DP universality against detailed vortex statistics, e.g., in the vortex-gas construction.

\begin{figure}

\includegraphics[width=1\linewidth,keepaspectratio]
{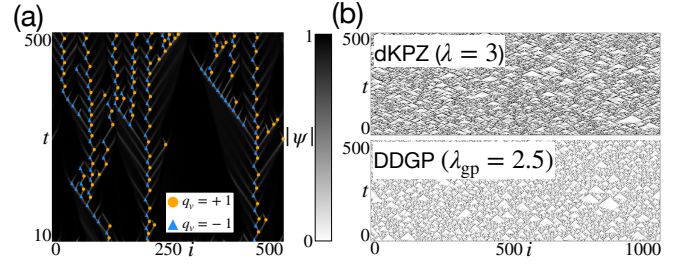}
\centering
\caption{Spacetime profiles of (a) the amplitude $|\psi|$ and associated vortices in the DDGP model at $\lambda_{\mathrm{gp}}=1.4$,
and (b) the order parameter $\rho$, for the discrete KPZ model at $\lambda=3$ (top) and the DDGP model at $\lambda_{\mathrm{gp}}=2.5$ (bottom).\label{fig:cgp}}

\end{figure}

\textit{\color{red}{Effective temporal discreteness from continuous-time dynamics.--}} Thus far, we have focused on vortices generated by discrete-time dynamics, such as Eq.~\eqref{eq:eom_phase}. As shown above, temporal discretization is essential for retaining vortices in a phase-only description. Yet fundamentally discrete-time evolution is uncommon in physical systems. Nevertheless, effective temporal discreteness can emerge from amplitude fluctuations: where the amplitude vanishes, the phase becomes ill-defined, appearing as a discontinuous phase jump in a phase-only description. We verify this mechanism numerically in a continuous-time model, i.e., the driven-dissipative Gross-Pitaevskii (DDGP) equation~\cite{carusotto2013rmp},
\begin{equation}
\partial_{t}\psi=D_{\text{gp}}\ [(\mathcal{D}_x)^{2}\psi]+\left(1+i\lambda_{\text{gp}}\right)\left(1-\left|\psi\right|^{2}\right)\psi.\label{eq:cGP}
\end{equation}
Likewise, this model exhibits a transition driven by vortex proliferation.
Figure~\ref{fig:cgp}(a) shows an exemplary spacetime snapshot of
$|\psi|$ and associated vortices for $D_{\mathrm{gp}}=1/2$ and $\lambda_{\mathrm{gp}}=1.4$, while Fig.~\ref{fig:cgp}(b) highlights the close resemblance between the $\rho$ profiles of this model and the discrete KPZ model. Moreover, the critical exponents agree with DP (Appendix~\hyperref[sec:cGP_numerics]{D}), supporting a continuous transition. Earlier work on a stochastic model~\cite{he2017prl} numerically inferred a first-order transition, but based on an order parameter not fully suitable to distinguish a weak first-order from a continuous one. 
Finally, at fixed finite $\lambda_{\mathrm{gp}}$, we observe long-lived particles at large $D_{\mathrm{gp}}$, which may drive the transition away from DP~\cite{bohr2001prl}, a possibility left for future work.

\textit{\color{red}{Conclusion and outlook.--}}
Our work uncovers a new far-from-equilibrium route to universal critical behavior. 
As in the BKT transitions, criticality is driven by defect proliferation. There, defects arise from equilibrium thermal or quantum fluctuations, with a bare fugacity set by their core action. Here, by contrast, they are generated by the nonequilibrium dynamics, tying their nucleation to local phase inhomogeneity. The resulting feedback produces effective multiplicative noise and drives DP criticality despite deterministic microscopic dynamics. While DP has previously been identified in deterministic systems~\cite{chate1987prl,grassberger1991pd,janaki2003pre,avila2023arfm,lemoult2024np,deger2022prl} and in transitions between topologically different turbulent states~\cite{takeuchi2007prl}, our results establish a distinct microscopic mechanism based solely on symmetry and topology, in which topological events themselves generate the emergent DP dynamics.
More broadly, they point to a new arena for universality and to mechanisms of criticality with no equilibrium counterpart, arising from the rich nonequilibrium dynamics of topological defects, e.g., self-propelled and unbinding disclinations in active nematics~\cite{giomi2013prl,giomi2014ptrs, shankar2018prl}, motile and fissioning dislocations in nonreciprocal solids~\cite{guillet2025pnas}, and spiral defects in nonreciprocal spin systems~\cite{avni2025prl}. Exciton-polariton platforms provide a timely setting in which the associated spatiotemporal critical scaling could be probed through coherence functions~\cite{fontaine2022nature,widman2026science}.

\begin{acknowledgments}
\textit{\color{red}{Acknowledgments.--}} We thank Ehud Altman, Siddhartha Dam, Haye Hinrichsen, Sven H\"ofling, Sebastian Klembt,   Leo Radzihovsky, Foster Thompson, Su-Chan Park, Yoshito Watanabe, Simon Widmann and Carl Zelle for discussions, and Yoshito in particular for assistance with numerics. 
Z.-M.~H., G.~J. and S.~D. are supported by the  Deutsche Forschungsgemeinschaft (DFG, German Research Foundation) under Germany's Excellence Strategy Cluster of Excellence Matter and Light for Quantum Computing (ML4Q) EXC 2004/1 390534769, and G.~J. and S.~D. by the DFG Collaborative Research Center (CRC) 1238 project number 277146847.
Numerical simulations were performed using \texttt{DifferentialEquations.jl}~\cite{rackauckas2017jors}, and executed on the RAMSES cluster at RRZK Cologne.
\end{acknowledgments}

\appendix
\clearpage
\clearpage
\section*{End Matter}

\section{Appendix A: Linear stability analysis of Eq.~\eqref{eq:dKPZ} }\label{sec:Linear_stability_anlysis}

We analyze the linear stability of Eq.~\eqref{eq:dKPZ} about a uniform-gradient
solution $u_{0}$. As shown in Eq.~\eqref{app_eq: linear_stab} below, the
$\lambda\to0$ limit yields the stability condition, 
\begin{equation}\label{app_eq:linear_stab_lim}
0\leq\left(2Ddt\right)\times\cos u_{0}\leq1.
\end{equation}
For convenience, we work with the $u\left(i,t\right)$ field, whose
dynamics obeys
\begin{equation}
\mathcal{D}_{t}u\left(i,t\right)=\mathcal{D}_{x}\mathcal{K}_{i}\left(u\right)=\mathcal{K}_{i+1}\left(u\right)-\mathcal{K}_{i}\left(u\right),
\end{equation}
or equivalently, 
\begin{equation}
u\left(i,t+dt\right)=u\left(i,t\right)+\left[\mathcal{K}_{i+1}\left(u\right)-\mathcal{K}_{i}\left(u\right)\right]dt.
\end{equation}
We then examine the stability by perturbing a uniform solution $u_{0}$
with $\delta u\left(i,t\right)$ and keeping terms to linear order
in $\delta u$, i.e., 
\begin{equation}
u\left(i,t\right)=u_{0}+\delta u\left(i,t\right),\ \delta u\left(i,t\right)\equiv\delta u\left(k,t\right)e^{ikx_{i}}.
\end{equation}
Here, the spatial coordinate in the exponent is written as $x_{i}$
to avoid confusion with the imaginary unit $i$. Correspondingly,
the Fourier amplitudes then evolve according to 
\[
\delta u\left(k,t+dt\right)=G\left(k\right)\delta u\left(k,t\right),
\]
 with the amplification factor, 
\begin{equation}
G\left(k\right)=1+\left(-4D\cos u_{0}\sin^{2}\frac{k}{2}+i\lambda\sin u_{0}\sin k\right)dt.
\end{equation}
Consequently, linear stability requires that 
\begin{equation}
\left|G\left(k\right)\right|\leq1,\ \forall k,
\end{equation}
or explicitly, 
\begin{equation}
\left(1-4D\cos u_{0}\sin^{2}\frac{k}{2}dt\right)^{2}+\left(\lambda\sin u_{0}\sin kdt\right)^{2}\leq1.\label{app_eq: linear_stab}
\end{equation}

In the $\lambda\to0$ limit, Eq.~\eqref{app_eq: linear_stab} reduces
to 
\[
0\leq2D\cos u_{0}\sin^{2}\frac{k}{2}dt\leq1,
\]
so requiring this inequality for all $k$ reproduces Eq.~\eqref{app_eq:linear_stab_lim}

\begin{figure}[htbp!]
\includegraphics[width=1\linewidth,
        keepaspectratio]{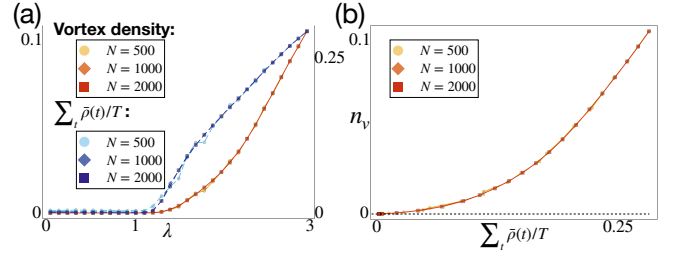}

\caption{Numerical results for (a) the vortex density  and spacetime-averaged $\rho$ as functions of $\lambda$, and (b) the vortex density as a function of the spacetime-averaged $\rho$. \label{app_fig:vortex}}
\end{figure}

\section{Appendix B: Numerical results for spacetime-averaged $\rho$ and the spacetime vortex density}
\label{sec:rho_vortex}
We numerically compute the vortex density ($n_v$) and the spacetime-averaged $\rho$ (i.e., $\frac{1}{T}\sum_t\bar\rho$) for several values of $\lambda$ and system sizes $N$ in the discrete KPZ model, using $D=1/2$, $dt=1/4$, a maximum simulation time of $2N$, and $100$ trajectories with random initial conditions, see Figure~\ref{app_fig:vortex}. In particular, 
panel (a) shows that the transition is accompanied by proliferation of spacetime vortices. Meanwhile,
panel (b) shows that the vortex density vanishes when the spacetime-averaged $\rho$ is zero (i.e., $n_v\propto \frac{1}{T}\sum_t\bar\rho$ for small $\rho$), providing numerical evidence for the absence of vortices in a homogeneous phase field.

\section{Appendix C: Numerical results for the continuous-time KPZ model}
\label{sec:cKPZ_numerics}

\begin{figure}[!htbp]
    \centering
\includegraphics[width=1\linewidth,
        keepaspectratio]{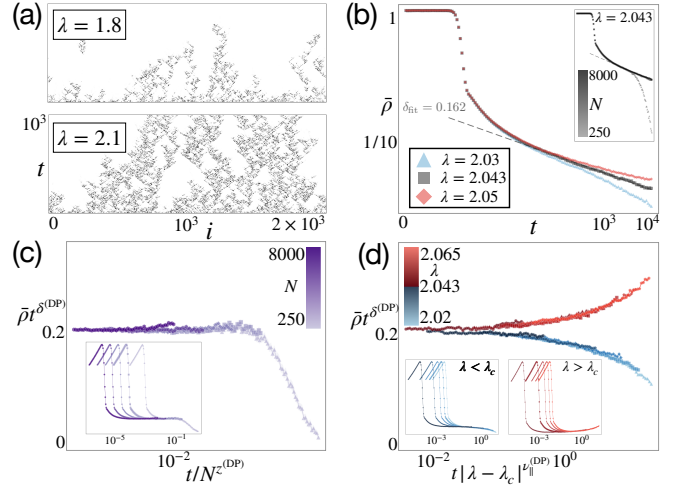}

\caption{Numerical results for the continuous-time limit of the discrete KPZ
model with $D=1/2$. As a stringent test, we choose the noise $\langle\xi\left(i,t\right)\xi\left(i^{\prime},t^{\prime}\right)\rangle=\sin^{2}\left[u(i,t)\right]\delta(t-t')\delta_{i,i'}$: Its variance vanishes at the targeted configuration $u=0$, and also at the additional $u=\pi$ configuration. Nevertheless, the resulting critical scaling remains consistent with DP. 
The results are averaged over 500 trajectories with fully activated
initial conditions [i.e., $u\left(i,t\right)=\pi+10^{-5}$]. (a) Spacetime
profile of $\rho(i, t)$. (b)-(d) Critical
scaling analyses showing agreement with the DP critical
exponents. The data-collapse analyses are performed over the time window $t\in[200,9000]$.
The numerical results locate the critical point at $\lambda_{c}\simeq2.043$.
Also, panel (d) is obtained for a system size $N=4000$. \label{app_fig:cKPZ}}

\end{figure}
Here, we show numerically that the continuous-time model,
\begin{equation}
\begin{cases}
\partial_{t}u\left(i,t\right)=\mathcal{D}_{x}\mathcal{K}_{i}+\xi\left(i,t\right)\\
\langle\xi\left(i,t\right)\xi\left(i^{\prime},\ t^{\prime}\right)\rangle=8\pi^{2}\kappa\left(u\right)\delta\left(t-t^{\prime}\right)\delta_{i,i^{\prime}}
\end{cases},
\end{equation}
exhibits critical scaling consistent with the DP universality class (see Fig.~\ref{app_fig:cKPZ}).

\section{Appendix D: Numerical results for the driven-dissipative Gross-Pitaevskii model [Eq.~\eqref{eq:cGP}]}
\label{sec:cGP_numerics}

\begin{figure}[h!]
\includegraphics[width=1\linewidth,
        keepaspectratio]{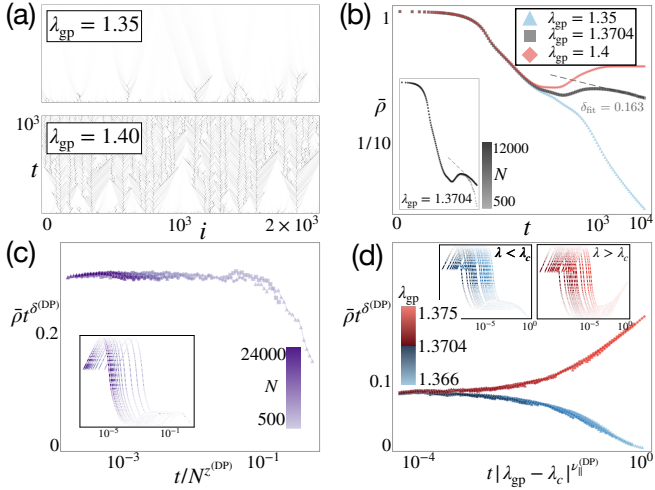}

\caption{Numerical results for the driven-dissipative Gross-Pitaevskii model with $D_{\mathrm{gp}}=1/2$,
averaged over 500 trajectories, initialized with random phases and unit amplitude. (a)
Spacetime profiles of $\rho(i, t)$. (b) Critical decay for a
system of size $N=8000$, shown on a log-log scale. At the estimated critical point, $\lambda_{c}\simeq1.3704$,
the fitted decay exponent is $\delta_{\mathrm{fit}}=0.163$. (c),
(d) Data-collapse plots using the DP critical exponents
over the time window $t\in[1500,10000]$. \label{app_fig:cGP}}
\end{figure}

As in the continuous-time phase model discussed above, we numerically
demonstrate that the driven-dissipative Gross-Pitaevskii model exhibits DP
criticality; see Fig.~\ref{app_fig:cGP}.

\end{document}